\documentclass[twocolumn,amsmath,amssymb]{revtex4}
\def\gsim{ \lower .75ex \hbox{$\sim$} \llap{\raise .27ex \hbox{$>$}} }
\def\lsim{ \lower .75ex \hbox{$\sim$} \llap{\raise .27ex \hbox{$<$}} }

\usepackage{graphicx}
\usepackage{dcolumn}
\usepackage{bm}
\input epsf

\newcommand{\be}{\begin{equation}}
\newcommand{\ee}{\end{equation}}
\newcommand{\bea}{\begin{eqnarray}}
\newcommand{\eea}{\end{eqnarray}}

\def\k{\kappa}

\begin{document}

\title{{\it Planck2013} results support the cyclic universe}
\author{Jean-Luc Lehners}
\affiliation{Max-Planck-Institute for Gravitational Physics
(Albert-Einstein-Institute), 14476 Potsdam, Germany}
\author{Paul J. Steinhardt}
\affiliation{Harvard-Smithsonian Center for Astrophysics, Harvard University, Cambridge, MA 02138 USA \\ Princeton Center for Theoretical Science,
Princeton University, Princeton, NJ 08544 USA \\ Joseph Henry
Laboratory, Jadwin Hall, Princeton University, Princeton, NJ 08544 USA}

\begin{abstract}
We show that results from the {\it Planck} satellite reported in 2013 are consistent with cyclic models of the universe for
natural parameter ranges ({\it i.e.}, order unity dimensionless coefficients), assuming the standard entropic mechanism for generating curvature perturbations.  With improved precision, forthcoming results from {\it Planck} and other experiments should be able to test the remaining parameter range and confirm or refute the core predictions - {\it i.e.}, no observable primordial B-mode polarization and detectable local non-gaussianity. A new prediction, given the {\it Planck2013} constraints on the bispectrum,  is a sharp constraint on  the local trispectrum parameter $g_{NL}$; namely, the currently best-understood models predict it is negative, with  $ g_{NL} \lesssim -1700$.
\end{abstract}

\maketitle

The recent results by the {\it Planck} satellite have been presented as supporting the predictions of some of the earliest and simplest
inflationary models of the early universe
 \cite{Ade:2013lta,Ade:2013rta,Ade:2013ydc}, which can all be reformulated as single-field models with a plateau type potential \cite{Ijjas:2013vea}. However, these models have a severe initial conditions problem  and typically lead to
eternal inflation, so that, in claiming to match the 'predictions' of these models, one is
implicitly making a number of unstated and poorly-understood assumptions about measures
and the multiverse \cite{Ijjas:2013vea}. This is not an insignificant point: with even moderately different
initial conditions and/or with some of the simplest measures (such as physical volume
weighting) the predictions of these simplest inflationary models are in fact in complete
{\it{disagreement}} with the {\it Planck2013} results.  Finally, within the context of the inflationary paradigm, the observationally favored potentials are highly unlikely since they are less generic, require more tuning, inflate for a smaller range of field values, and produce exponentially less inflation than the power-law potentials \cite{Ijjas:2013vea}.  Thus, looking for alternatives to inflation remains as well motivated
as ever.

In particular, the cyclic theory of the universe \cite{Steinhardt:2002ih} does not appear to suffer from these same problems.  In the present paper, we demonstrate that
 cyclic models can match all of the cosmological parameters
favored by {\it Planck}. These models are characterized by simple potentials;
by natural parameter ranges ({\it i.e.}, with dimensionless quantities of order unity); and they assume the standard entropic mechanism for generating curvature perturbations \cite{Lehners:2007ac}.  At the same time, forthcoming results from {\it Planck} and other experiments will reach the point where they should be able to explore the natural ranges and confirm or refute the core predictions - namely, no observable primordial B-mode polarization and detectable local non-gaussianity in the bispectrum.  Furthermore, given the {\it Planck2013} constraints on the bispectrum, we present a new test, namely a sharper prediction for the local trispectrum parameter $g_{NL}$.

Our conclusions differ markedly from those reported by the {\it Planck2013} collaboration who describe the ekpyrotic/cyclic scenarios as being under ``severe pressure'' \cite{Ade:2013ydc} as a result of the new data.  We note that the basis of their summary is an analysis of a particular form of scalar field potential (exponential); a particular mechanism of generating curvature perturbations (entropic, as in this paper); and, within that mechanism, a particular choice of conversion from entropic to curvature perturbations occurring during the ekpyrotic smoothing phase \cite{Buchbinder:2007at}, which, in the cyclic model, is far less likely than the alternative, a conversion after the smoothing phase is over -- see \cite{Lehners:2011ig}. A fair analogue would be to judge the inflationary scenario on the basis of how well an exponential inflaton potential fits the {\it Planck2013} data: the answer would be it is ruled out at more than the 99.7\% confidence level \cite{Ade:2013rta}. Obviously, it would be wrong to rule out a whole scenario on the basis of a single type of potential.  Furthermore, in their detailed discussion, the {\it Planck2013} collaboration reports that the same data {\it fits} the cyclic predictions for a range of parameters just by shifting the conversion to the kinetic phase occurring after the smoothing phase and still keeping to the strict exponential form \cite{Ade:2013ydc}.  This {\it Planck2013} result is consistent with our analysis below, which assesses the situation introducing a more model-independent parameterization of the ekpyrotic potential that enables us to quantify the fit in an unbiased way.  At the same time, we note that a new prediction regarding the trispectrum resulting from the Planck measurements in the context of the entropic mechanism.  More generally, given the serious and exciting scientific issues at stake, such as the origin and evolution of the universe, the nature of the cosmic singularities, and the existence of a multiverse versus universe, we hope data analysts will take some care not to reach conclusions about a cosmological scenario prematurely.

\noindent
{\it Cyclic Model.}  The cyclic model \cite{Steinhardt:2001st,Lehners:2008vx} is based on the idea that the big bang is a big bounce from a phase of contraction to a phase of expansion; that the key events that set the large-scale smoothness, flatness and density perturbations occurred before the bounce during a phase of dark energy domination (similar to today) and ultra-slow ({\it ekpyrotic}) contraction with equation of state $w \gg 1$ \cite{Khoury:2001wf} (with no high energy inflation after the bounce); and that the bounces regularly repeat such that phases of expansion and contraction alternate.  Unlike earlier oscillatory models, there is a strong asymmetry between phases, with each cycle undergoing
vastly more expansion than contraction. Thus, during each cycle the scale factor $a$
grows by a huge factor. Nevertheless, local quantities, such as the Hubble parameter $H
\equiv \dot{a}/a,$ temperature, and density return to the same value after each cycle, and it is in this sense that
the models are cyclic.

The ekpyrotic phase can be generated by
 a scalar field $\sigma$ evolving down a steep and negative potential
\begin{equation}
V = -V_0 e^{\sqrt{2 \epsilon}\sigma} [1 + \cdots],
\end{equation}
where $V_0$ is a constant, $\epsilon = \frac{3}{2}(1+w)$ is in general a slowly-varying function of time and
the ellipsis stands for additional terms that parameterize the potential in terms of a second
scalar field $s$ that we will discuss below.

The big bounce has been modeled in two inequivalent ways so far: in the first class of models the bounce is classically non-singular in the sense that the scale factor $a$ reverses from contraction to expansion before reaching zero. Such models typically rely on higher-derivative kinetic terms \cite{Buchbinder:2007ad,Qiu:2011cy,Easson:2011zy,Cai:2012va,Osipov:2013ssa,Qiu:2013eoa}. In the second class of models, which were originally motivated by the braneworld picture suggested by string theory, the bounce corresponds to the collision of orbifold branes. In this case, the evolution is classically singular ($a$ reaches zero), but a semi-classical treatment indicates that this singularity is resolved at the quantum level \cite{Turok:2004gb}. In either case, the bounce is an analytic and unitary event so that all physical quantities vary continuously across the bounce, which is all that needs to be known to make predictions.

Notably, the Hubble rate is largest right at the beginning of the radiation phase after the bounce, and is significantly smaller during the 'smoothing' dark energy and ekpyrotic phases. Thus, quantum fluctuations during the smoothing phases are always small, and consequently no run-away behavior, such as the one leading to eternal inflation, occurs in these models \cite{Johnson:2011aa,Lehners:2012wz}. This feature makes it possible to avoid the unpredictability problem of inflation.

{\it Density perturbations.} The best-understood way in which nearly scale-invariant curvature perturbations can be
generated during a contracting phase of the universe is via the entropic mechanism \cite{Finelli:2002we,Lehners:2007ac}. In
this scenario, there is a second scalar field $s,$ which develops nearly scale-invariant
entropy (or isocurvature) perturbations, and these entropy perturbations are subsequently
converted into curvature perturbations during the kinetic phase just before the bounce.

We will parameterize the potential for both scalar fields during the ekpyrotic phase as
\begin{equation}
V = -V_0 e^{\sqrt{2 \epsilon}\sigma} [1 + \kappa_2 \epsilon s^2 + \frac{\kappa_3}{3!}
\epsilon^{3/2} s^3 + \frac{\kappa_4}{4!} \epsilon^2 s^4 +\cdots ],
\end{equation}
where $\kappa_2,\kappa_3,\kappa_4$ are constants. A potential of this form can for
example be constructed by having two scalar fields $\phi_1,\phi_2$ with separate
ekpyrotic potentials $V=-V_1 e^{c_1 \phi_1}-V_2 e^{c_2 \phi_2}$ and then performing
a rotation in field space into the ekpyrotic direction $\sigma$ (defined to point tangentially to the background trajectory, with $\dot{\sigma}=(\dot{\phi}_1^2+\dot{\phi}_2^2)^{1/2}$) and the transverse direction $s$. In this case, the $\kappa_{2,3,4}$ parameters are
\begin{equation}
\kappa_2 = 1, \quad \kappa_3 = 2 \sqrt{2} \frac{(c_1^2 - c_2^2)}{|c_1 c_2|}, \quad
\kappa_4 = 4 \frac{(c_1^6 + c_2^6)}{c_1^2 c_2^2 (c_1^2 + c_2^2)}, \label{eq1}
\end{equation}
with $1/\epsilon = 2/c_1^2 + 2/c_2^2.$ These expressions allow us to estimate natural
parameter ranges: we may for example expect $\kappa_2$ to be close to $1,$ say within
$10\%$ -- see also a more detailed earlier discussion in \cite{Buchbinder:2007tw}. One may also expect that in a fundamental physics context, the values of $c_1$
and $c_2$ that one can obtain are not wildly different. In this case the potential is roughly
symmetric across $s=0,$ and as we will see below, this case is the one of most interest.
Then we may expect $\kappa_3 \in [-1,1]$ and $\kappa_4 \approx 4.$  The ekpyrotic phase ends when the steeply falling potential bottoms out at $V_{ek-end}$, a short time before the bounce.  We are now ready
to turn to a calculation of the linear fluctuations in these models.

There are two gauge-invariant scalar fluctuations to consider: the curvature fluctuation
$\zeta,$ defined as a perturbation to the scale factor in a flat FLRW universe,
\begin{equation}
ds^2 = -dt^2 + a^2(t)e^{2\zeta(t,x^i)}dx^i dx_i,
\end{equation}
and the entropy fluctuation $\delta s.$ During
the ekpyrotic phase, the curvature perturbations obtain a very blue spectrum with a small
amplitude, and hence can be neglected \cite{Lyth:2001pf}. However, the entropy perturbations are of more
interest. They obey the equation of motion
\begin{equation}
\ddot{\delta s} + 3 H \dot{\delta s} + V_{,ss} = 0.
\end{equation}
A standard calculation then shows that this equation can be solved in terms of Hankel functions, and the boundary condition that in the far past the solution should approximate the Minkowski vacuum fixes the solution to be given by $\delta s = \frac{\sqrt{-\pi t}}{2}H^{(1)}_\nu (-kt),$ where
the Hankel function has index
$\nu = (\frac{1}{4}+2\kappa_2 - \frac{2 \kappa_2}{\epsilon} - \frac{1}{\epsilon}+
\frac{3}{2}\kappa_2
\frac{\epsilon_{,{\cal N}}}{\epsilon})^{1/2}.$ Here ${\cal N}$ denotes the number of e-folds
left before the end of the ekpyrotic phase ($d{\cal N} = d(\ln aH)$) and terms of order $1/\epsilon^2$ have been neglected.
At late times $(-kt)
\rightarrow 0$ we obtain $\delta s \approx
\frac{1}{\sqrt{2}(-t)k^\nu},$ implying that at the end of the
ekpyrotic phase, the entropy perturbation is given by \begin{equation} \delta
s(t_{ek-end}) \approx \frac{|\epsilon V_{ek-end}|^{1/2}}{\sqrt{2}k^\nu}.
\end{equation}
Hence, the spectral index is given by \begin{equation} n_s -1 = 3-2\nu \approx \frac{4}{3}(1 - \kappa_2) + \frac{2}{\epsilon } -
\frac{\epsilon_{,{\cal N}}}{\epsilon}, \label{spectrum}\end{equation} where the last
approximation is valid as long as $\kappa_2$ is close to $1$, {\it i.e.} as long as
$|\kappa_2 -1| \ll 1.$ Thus, the spectrum typically deviates from exact scale-invariance by
a few percent, either to the blue or to the red. The term in $\kappa_2$ can by itself render
the spectrum red or blue. The $2/\epsilon$ contribution shifts the spectrum to the blue,
while the $\epsilon_{,{\cal N}}/\epsilon$ term tends to shift it to the red, as
towards the end of the ekpyrotic phase $\epsilon$ can be expected to decrease while
${\cal N}$ decreases by definition. Thus, the entropic mechanism does not yield a clear
preference for either a red or a blue tilt. However, for natural values of the parameters,
the spectrum can be in perfect agreement with the measured {\it Planck2013} value
$n_{s} = 0.9603 \pm 0.0073$ \cite{Ade:2013lta}, for example by having $\kappa_2 = 1.06$ and
constant $\epsilon = 50.$ (Recall that $\epsilon$ is typically of order $1/{\cal N}$ in inflationary models and of order ${\cal N}$ in ekpyrotic/cyclic models where ${\cal N} \sim 50$ for the modes of observable interest in both cases.)

Conversion of the  entropy perturbations into curvature
occurs generically when the trajectory in scalar field space bends (for a concrete realization in supergravity, see \cite{Lehners:2006pu,Lehners:2006ir}).
The result of this bending is that a (linear) curvature perturbation $\zeta_L$ develops,
with a spectrum identical to that of the entropy perturbation $\delta s$ and with an
amplitude
$\zeta_L \approx \frac{1}{5} \delta s(t_{ek-end})$.
Thus, the amplitude of the resulting curvature perturbations is given by
\begin{equation}
\langle \zeta_L^2 \rangle \approx \int \frac{dk}{k}\frac{\epsilon V_{ek-end}}{10^3},
\end{equation}
so that the measured variance
can be achieved for ekpyrotic potentials bottoming out at $V_{ek-end}$ around the grand unified
scale \cite{Lehners:2010ug}.

{\it Non-Gaussianities.}  Aside from improving measurements of the power spectrum, a great advance in {\it Planck2013} is a substantially better bound on non-gaussianity \cite{Ade:2013ydc}.
Since any non-gaussian signals are seen as clear indicators for something
other than simple inflationary models, the results rule out a large range of complex inflationary models and other alternatives.  However, we show below that the current limits fit simple cyclic models for a substantial natural range of parameters.

For cyclic models, one intuitively expects significant departures from gaussianity compared to inflation because the requisite potential is steeper and thus self-interactions of the scalar fields are
guaranteed \cite{Creminelli:2007aq,Buchbinder:2007at,Koyama:2007if,Lehners:2010fy}. But, as we will review below, this does not imply that every non-gaussian
estimator must be large. 
In current models, the kinetic terms for
the scalar fields are of canonical form. Hence, the non-gaussian corrections that are
generated are of the local form, and they can be analyzed by studying the classical
equations of motion to higher orders in perturbation theory. It
is useful to define the following expansion for the curvature perturbation (on uniform
energy density hypersurfaces)
\begin{equation}
\zeta= \zeta_L + \frac{3}{5}f_{NL} \zeta_L^2 + \frac{9}{25}g_{NL}
\zeta_L^3,
\end{equation}
where the numerical coefficients are conventional. The parameters $f_{NL}$ and
$g_{NL}$ then describe the deviations from gaussianity of the bispectrum and
trispectrum respectively.

In order to calculate these parameters, we first have to work out the entropy perturbations
to third order in perturbation theory. A straightforward, but lengthy calculation yields \cite{Lehners:2009ja}
\be
\delta s= \delta s_L + \frac{\kappa_3 \sqrt{\epsilon}}{8}\delta s_L^2 +
\epsilon(\frac{\kappa_4}{60}+\frac{\kappa_3^2}{80}-\frac{2}{5})\delta s_L^3,
\end{equation}
where terms of order $1/\epsilon$ have been neglected. In order to calculate how this entropy perturbation gets converted into a non-linear
curvature perturbation, it is convenient to use the formula \cite{Lehners:2009qu}
\begin{equation}
\dot\zeta= \frac{2 \bar H
\delta V}{\dot{\bar \sigma}^{2} -2 \delta V},
\end{equation}
which is valid to all orders and where an overbar denotes a background quantity. One can
use this simple formula to derive an analytical estimate of the non-linear curvature
perturbation. However, a more precise procedure is to integrate the above formula numerically. This
was done in \cite{Lehners:2007wc,Lehners:2008my,Lehners:2009ja} for several different implementations of the bending of the
background trajectory. There it was found that the results are most robust when the
conversion process is relatively smooth ({\it i.e.} lasting on the order of one Hubble time -
shorter conversion times tend to amplify the magnitudes of the non-linearities somewhat),
and we will focus on this case. The resulting non-gaussianity parameters can be well
fitted by the approximate formulae
\begin{eqnarray}
f_{NL} &=& \pm 5 + \frac{3}{2}\kappa_3\, \sqrt{\epsilon} \label{fNL}\\
g_{NL} &=& (-40+\frac{5}{3}\k_4+\frac{5}{4}\k_3^2)\, \epsilon,\label{gNL}
\end{eqnarray}
which also agree well with the analytical estimates.

Along with the spectral tilt in Eq.~(\ref{spectrum}), these two expressions comprise the key predictions of the scalar density perturbations (generated via the entropic mechanism) to compare with {\it Planck2013} and future observations.  In addition, ekpyrotic/cyclic models predict no observable tensor modes on large scales \cite{Boyle:2003km,Baumann:2007zm}.

The expressions here differ in form from previous work because we have introduced the three dimensionless model-independent parameters $\kappa_i$ to judge the natural range expected in this cyclic model. The essential result remains: the model predicts substantially more non-gaussianity than simple inflationary models.  For example, whereas the total $f_{NL}$ for simple inflation models is less than one \cite{Maldacena:2002vr}, we see that the formula for $f_{NL}$ in the cyclic model contains a term $\pm 5,$ which arises from the conversion
process. Its sign depends on the details of the
conversion process ({\it e.g.} the trajectory bending to the left rather than to the right in field space).  The second term in Eq.~({\ref{fNL}) is also substantially greater than one.
However, allowing for either possibility ({\it i.e.} plus or minus sign), we find that the {\it Planck}
bounds of $f_{NL} = 2.7 \pm 5.8$ at the $1 \sigma$ level can be matched respectively
by
\begin{equation}
-0.76 < \kappa_3 < 0.33 \quad \textrm{or} \quad 0.18 < \kappa_3 < 1.27
\end{equation}
where we have used again a typical value of $\epsilon = 50.$ These allowed ranges for $\kappa_3$ at the
$1 \sigma$ level are still of ${\cal O}(1),$ and thus, contrary to the claim in
\cite{Ade:2013ydc}, do not constitute ``severe pressure'' on cyclic models.  The fit remains good over a substantial range of $\epsilon$.  

In fact, fine-tuning is not required to achieve any value in the entire range of $f_{NL}$ allowed by {\it Planck2013}, including $f_{NL}$ near zero (as expected for simple inflation models).  Hence, observing $|f_{NL}|$ measurably greater than one would fit the simplest cyclic picture and be inconsistent with simple inflation models; but observing  $|f_{NL}|$ less than one would, taken by itself, not be conclusive.  

Fortunately, there is a another test to consider: the trispectrum. In addition to the parameter $g_{NL}$
that we are about to discuss, there exists a second trispectrum parameter, denoted
$\tau_{NL},$ which for models where the perturbations originate from a single
perturbation mode (such as the entropy perturbation here) takes the form (see {\it e.g.} \cite{Byrnes:2006vq}):
\begin{equation}
\tau_{NL} = \frac{36}{25}f_{NL}^2.
\end{equation}
Thus, given the bounds on $f_{NL}$ that we just described, a consistency check for
ekpyrotic models is that $\tau_{NL}$ must be smaller than about $100$ at the $1
\sigma$ level. The constraints from the {\it Planck} satellite for this parameter are
$\tau_{NL}<2800$ (95\% confidence level), and thus these are consistent.

Less well-constrained, but probably also more interesting, is the intrinsic local
trispectrum parameter $g_{NL}.$ The strong bounds on $f_{NL}$ imply that
$\kappa_3$ cannot be much greater than unity, and in turn this implies that $\kappa_4
\approx 4$ (see the discussion below Eq. (\ref{eq1})). Using these relations, we can
simplify the fitting formula for the trispectrum, so that it ends up depending solely on the
equation of state parameter $\epsilon,$
\begin{equation}
g_{NL} \approx - 35 \, \epsilon.
\end{equation}
This remarkably simple formula constitutes a key test for the simple cyclic models described here. The percent level deviation of the spectral tilt from exact scale invariance is natural if $\epsilon$ is equal to or larger than about $50,$ {\it cf.} Eq. (\ref{spectrum}) -- this consideration immediately leads to an approximate upper bound on $g_{NL}$ of about $g_{NL} \lesssim -1700.$ A larger equation of state parameter $\epsilon$ then leads to a correspondingly larger magnitude of $g_{NL}$.

What is interesting is that {\it the entropic mechanism makes a definite prediction for the sign of
$g_{NL},$ namely that it should be negative.} We note that a larger $\epsilon$ allows a smaller range of $\kappa_3$ values to be consistent with the current bounds on $f_{NL}$ ({\it cf.} Eq. (\ref{fNL})), so that one may conclude that values of $g_{NL}$ closer to the upper bound, $ -1700$, are more natural than substantially larger negative values. The robustness of the result is due to the large contribution from the
$\kappa_3,\kappa_4$-independent term in Eq. (\ref{gNL}). Although current bounds on
$g_{NL}$ are still weak (they are of order $\pm 10^5$ \cite{Smidt:2010ra,Sekiguchi:2013hza}), this prediction is interesting for
future observations, as the predicted values are much larger in
magnitude than those of simple inflationary models.

\noindent
{\it Discussion:} The data recently released by the {\it Planck} satellite team is extraordinarily precise. It shows that the primordial perturbations were nearly scale-invariant, adiabatic and gaussian. These findings have immediately been hailed as confirming the predictions of some of the simplest models of inflation. However, as emphasized in Ref.~\cite{Ijjas:2013vea}, at the same time the data favors certain models, it disfavors the paradigm. The particular models favored by the data are a set that is exponentially unlikely to have produced our universe compared to more generic power-law potentials and only produce inflation in cases where the universe is already surprisingly smooth over many Hubble volumes before inflation begins.

By contrast, we have shown in this paper that a radically different class of models without these problems, namely cyclic models containing an ekpyrotic phase and using the standard entropic mechanism \cite{Lehners:2007ac}, match the {\it Planck} findings very well: the {\it Planck} central values of cosmic microwave parameters, as well as the entire $1 \sigma$ ranges 
, can be reproduced with natural $O(1)$ ranges of the parameters. At the same time, by constraining the spectral tilt and $f_{NL}$ to a significant degree, the current results imply that a new (falsifiable!) prediction for cyclic models is that the intrinsic trispectrum parameter $g_{NL}$ should be negative in sign and of order $10^3 - 10^4$ in magnitude.
(As a cautionary note, we point out that the method of generating curvature perturbations in cyclic models is, to a large degree, separate from the scenario overall, as in the case of inflation.  Other approaches than the one considered here would produce
different predictions for non-gaussianity.)
Whereas the non-gaussianity is somewhat model-dependent, a model-independent, falsifiable prediction of cyclic models is that there are no detectable primordial gravitational waves on the scales of cosmological interest.
Thus, near-future measurements will provide clear-cut tests of these simple cyclic models of the universe.

We would like to thank Bill Jones and David Spergel for reading the manuscript and for their valuable comments. JLL gratefully acknowledges the support of the European Research Council via the Starting Grant No. 256994. This work was supported in part by the US Department of Energy grant DE-FG02-91ER40671 (PJS).  PJS is grateful to the Simons Foundation and the Radcliffe Institute for providing support during his leave at Harvard and to the Institute for Theory and Computation at the Harvard-Smithsonian for hosting him during the period that this work was done.

\bibliography{PlanckCyclic}

\end{document}